\documentclass[prl,twocolumn,showpcs,amsmath,amssymb]{revtex4}
\usepackage[dvipdfm]{graphicx}% Include figure files 
\usepackage{dcolumn}% Align table columns on decimal point 
\usepackage{bm}% bold math \usepackage{amsmath} 

\begin{document}
%\bibliographystyle{h-elsevier}
%\draft

\author{Nasser Demir}
\affiliation{Department of Physics, Duke University, Durham, NC 27708, USA}
\author{Steffen A.~Bass}
\affiliation{Department of Physics, Duke University, Durham, NC 27708, USA}

\title{Shear-Viscosity to Entropy Density Ratio of a Relativistic Hadron Gas}

\date{\today}

\begin{abstract}
Ultrarelativistic heavy-ion collisions at the Relativistic Heavy-Ion Collider (RHIC) are thought to have produced a state of matter called the Quark-Gluon-Plasma, characterized by a very small shear viscosity to entropy density ratio $\eta/s$, near the lower bound predicted for that quantity by Anti-deSitter space/Conformal Field Theory (AdS/CFT) methods.  As the produced matter expands and cools, it evolves through a phase described by a hadron gas with rapidly increasing $\eta/s$. We calculate $\eta/s$ as a function of temperature in this phase and find that its value poses a challenge for viscous relativistic hydrodynamics, which requires small values of $\eta/s$ throughout the entire evolution of the reaction in order to successfully describe the collective flow observables at RHIC. We show that the inclusion of non-unit fugacities will reduce $\eta/s$ in the hadronic phase, yet not sufficiently to be compatible with viscous hydrodynamics.  We therefore conclude that the origin of the low viscosity matter at RHIC must be in the partonic phase of the reaction.
\end{abstract}

\maketitle

%\pagebreak
Ultrarelativistic heavy ion collisions at the Relativistic Heavy Ion Collider (RHIC) are thought to have produced a Quark Gluon Plasma (QGP) with the characteristics of a near ideal fluid\cite{Adcox:2004mh,Back:2004je,Arsene:2004fa,Adams:2005dq}. One of the most important current challenges
in QGP research is to quantify the transport coefficients of this novel state of matter.  Recently, attention in the field has been primarily focused on the shear viscosity to entropy density ratio $\eta/s$.  Certain supersymmetric gauge theories with gravity duals \cite{Policastro:2001yc} suggest a lower bound of $\eta_{\rm min} = s/4\pi$ for this quantity, often referred to as the KSS bound \cite{Kovtun:2004de}.  Relativistic viscous hydrodynamic calculations  require very low values of $\eta/s$ in order to reproduce the RHIC elliptic flow ($v_2$) data \cite{Song:2007fn,Romatschke:2007mq,Luzum:2008cw}.  However, current calculations assume a fixed value of $\eta/s$ throughout the entire evolution of the system and neglect its temperature dependence. The exact value of $\eta/s$ in these calculations is only known within a factor of $\approx 3$, due to systematic uncertainties related to the choice of equation of state and initial conditions used \cite{Luzum:2008cw,Kharzeev:2002ei}.
% the most prevalent Glauber initial condition assumes that the initial energy density in a heavy ion collision scales with the number density of binary collisions, whereas the Color-Glass-Condensate (CGC) initial conditions incorporate a scenario which relates the initial energy density to a gluon density predicted by the CGC model \cite{Kharzeev:2002ei,McLerran:1993ni,McLerran:1993ka}.
 A viscous hydrodynamical analysis \cite{Luzum:2008cw} finds that $\eta/s$ should lie within a range of 0.08-0.24 depending upon the choice of initial conditions and equation of state.  This finding is supported by calculations of $\eta/s$ for pure gluonic QCD, which yield values close to the KSS bound, and indirect estimates of $\eta/s$ from calculations of the diffusion of heavy quarks, elliptic flow measurements, and transverse momentum correlations that arrive at roughly comparable values \cite{Meyer:2007ic,Lacey:2006bc,Drescher:2007cd,Gavin:2006xd,Adare:2006nq}.  

Note that the shear viscosity of matter in a relativistic heavy ion collision is a \textit{time-dependent} quantity.  While the partonic phase of such a collision is expected to have a very low value of $\eta/s$, after hadronization occurs $\eta/s$ is expected to rapidly increase.  In order to quantitatively constrain the viscosity of the deconfined phase of a relativistic heavy ion collision, a separate calculation of the hadronic viscosity is necessary. Several investigations suggest $\eta/s$ should reach a  minimum in the vicinity of a phase transition or crossover \cite{Chen:2007jq,Csernai:2006zz}.  In particular, \cite{Csernai:2006zz} argue that $\eta/s$ should decrease as a function of rising temperature in the hadronic phase, and then increase in the deconfined phase.  They suggest that $\eta/s$ should reach a minimum near the deconfinement transition, but also remark that the perturbative methods used in their arguments are not applicable near $T_c$. It has also recently been argued that the existence of Hagedorn states \cite{NoronhaHostler:2008ju} will strongly decrease hadronic $\eta/s$, especially close to $T_c$.  It should be noted though that all calculations performed thus far have assumed kinetic and chemical equilibrium, equivalent to unity light quark fugacities.  Such an assumption may be reasonable for the formation of a QGP in a relativistic heavy ion collision at RHIC; yet the different timescales of chemical and kinetic freeze-out in the hadronic phase of the reaction imply an acquiring of non-unit particle species dependent fugacities  as the system evolves in the hadronic phase \cite{Bass:2000ib,Hirano:2002hv,kolb:2002ve,Nonaka:2006yn}.  

Several analytic calculations of $\eta$ and $\eta/s$ for simple hadronic systems have previously been performed \cite{Gavin:1985ph,Dobado:2001jf,Prakash:1993bt,Chen:2006iga,Chen:2007xe,Itakura:2007mx}.  These analytic calculations solved the linearized Boltzmann equation, in which the cross sections in the collision integral were treated using different techniques, such as chiral perturbation theory, effective NN theory, and phenomenological amplitudes.  However, even the most sophisticated analytic calculations include one or two hadronic species at most, and such a binary mixture clearly is a very crude approximation of the hadronic matter present at RHIC.  Sophisticated Monte Carlo microscopic transport models, which include the full range of the hadronic spectrum (including resonances) \cite{Bass:1998ca}, provide a far more realistic description of the hadronic matter created in relativistic heavy ion collisions. Several studies within those models have focused on equilibration and thermodynamic properties of infinite hadronic matter \cite{Belkacem:1998gy}, and have extracted transport coefficients of hadronic gases \cite{Muronga:2003tb,Muroya:2004pu}, albeit none have performed a systematic study of $\eta/s$.  

In this Letter, we use a microscopic transport model known as the Ultrarelativistic Quantum Molecular Dynamics (UrQMD) model, described in \cite{Bass:1998ca,Bleicher:1999xi}, to simulate infinite equilibriated hadronic matter.  We confine the particles comprising the system to a box with periodic boundary conditions in coordinate space \cite{Belkacem:1998gy}, and the collisions force the system into equilibrium.  We verify that the system has achieved chemical equilibrium by checking whether the particle multiplicities in our system saturate as a function of time, and comparing such yields to an independent statistical model of a hadron resonance gas (SHARE) \cite{Torrieri:2006xi}. We verify kinetic equilibrium by checking the momentum distributions of the hadrons in our system for isotropy, and fitting particle spectra to Boltzmann distributions. 
%We find that $\eta/s$ decreases at finite chemical potential(s) 
%in the hadronic phase of a relativistic heavy ion reaction, which is an important prerequisite for the understanding of the successful application of viscous
%fluid dynamics to elliptic flow data at RHIC.  We then conclude that the minimum $\eta/s$ must exist in the deconfined phase. 

\begin{figure}[tb]
\includegraphics[width=1.0\linewidth]{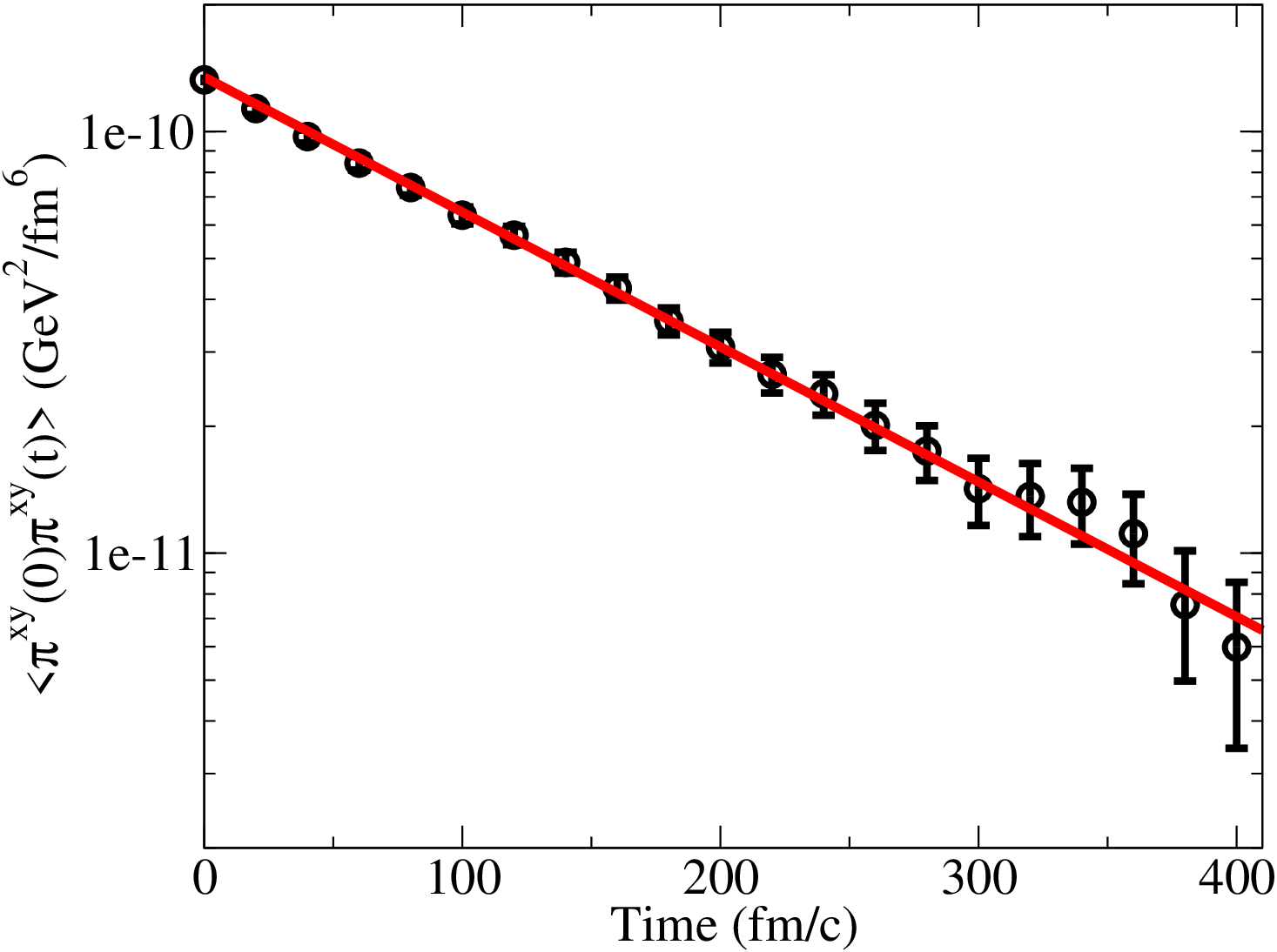}
\caption{The viscous correlator for T=67.9 MeV.  The relaxation time associated with this plot is 136 fm/c.}
\label{Fig-kubo_correlators}
\end{figure}

	In order to extract the shear viscosity of our system, we employ the Kubo formalism.  In addition, we compute the entropy of the system and evaluate $\eta/s$ as a function of temperature and baryo-chemical potential.  The Kubo formalism relates linear transport coefficients to near-equilibrium correlations of dissipative fluxes and treats dissipative fluxes as perturbations to local thermal equilibrium \cite{Hosoya:1983id,Paech:2006st,Kubo-formalism}.  The Green-Kubo formula for shear viscosity is 
\begin{equation}
\eta= \frac{1}{T} \int d^3r\int_0^{\infty}\,dt \langle \pi^{xy}(\vec{0},0)\pi^{xy}(\vec{r},t)\rangle_{\rm equil} ,
\end{equation}
where $T$ is the temperature of the system,  $t$ is the post-equilibriation time
 (the above formula defines \begin{math}t=0\end{math} as the time the system equilibriates), 
 and $\pi^{xy}$ is the shear component of the energy momentum tensor $\pi^{\mu \nu}$.  
 The expression for the energy momentum tensor $\pi^{\mu \nu}$ is 
\begin{equation}
\pi^{\mu \nu}= \int d^3p \frac{p^{\mu}p^{\nu}}{p^0} f(x,p),
\end{equation}where $f(x,p)$ is the phase space density of the particles in the system.  Our system does not assume any interparticle potential, and UrQMD treats the hadrons as point particles uniformly distributed in coordinate space, which implies
\begin{equation}
\pi^{xy}=\frac{1}{V}\sum_{j=1}^{\rm{N_{part}}}\frac{p^x(j)p^y(j)}{p^0(j)},
\end{equation} 
where $V$ is the volume of the system.	
In the Kubo formula, the averaging symbol denotes an averaging over the ensemble of events generated in our simulation.  A representative sample of the correlations of the shear component of the energy momentum tensor is given in Fig. 1.  The correlation functions are empirically found to decay exponentially in time, hence we assume an exponential ansatz to integrate the correlation function over time.     

\begin{figure}[tb]
\includegraphics[width=0.9\linewidth]{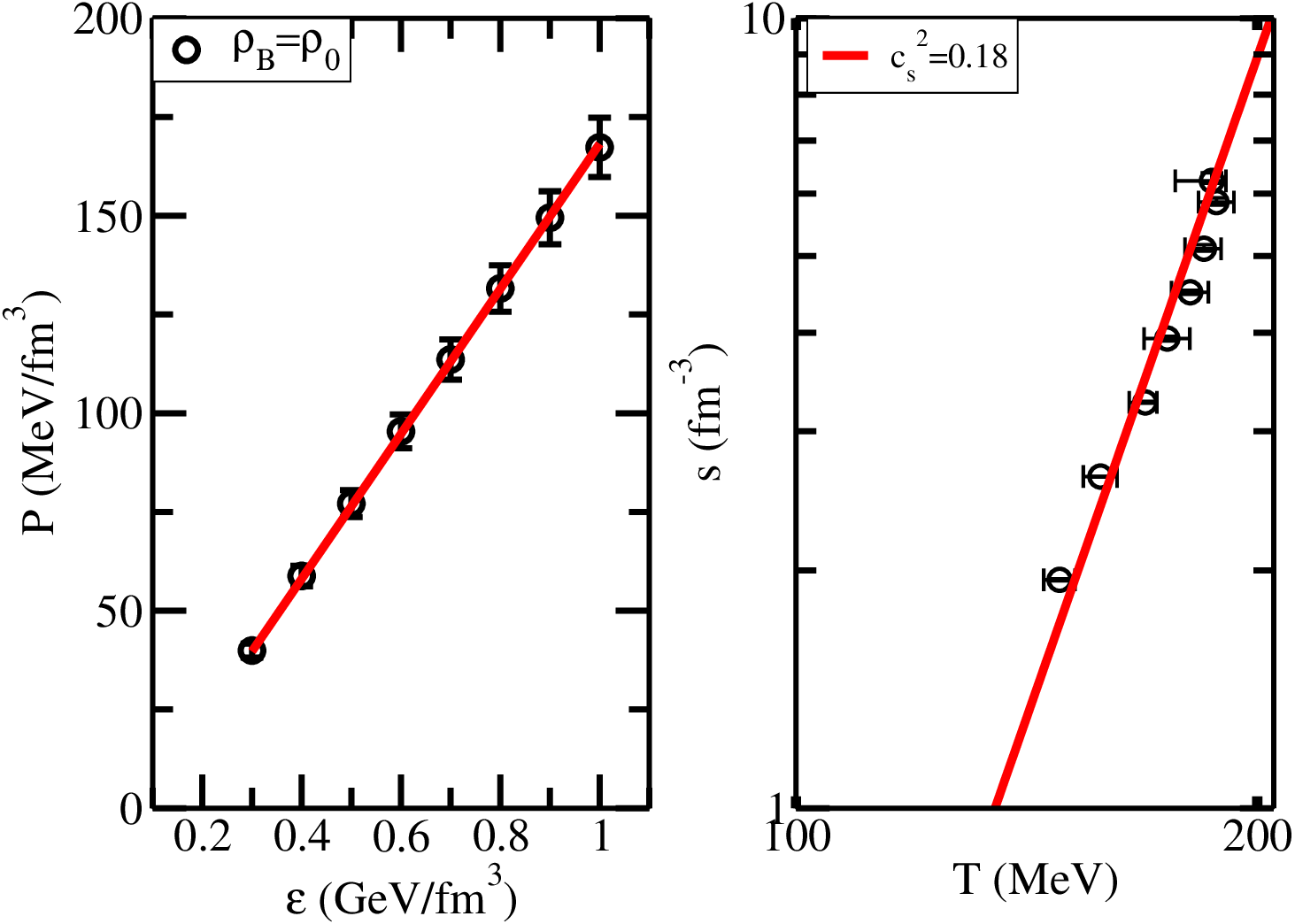}
\caption{In the left panel, pressure versus energy density.  The speed of sound extracted from the slope yields $c_s^2=0.18$.  In the right panel, entropy density versus temperature.}
\label{Fig-entropy_scaling}
\end{figure}

In order to compute the entropy of the system, we use the Gibbs 
formula $s=\left(\frac{\epsilon+P-\mu_B\rho_B}{T} \right)$.  The energy density, pressure, baryo-chemical potential, and number densities of the relevant chemical species are extracted once the system has equilibriated.  The pressure is computed via
\begin{equation}
P=\frac{1}{3V}\sum_{j=1}^{\rm{N_{part}}}\frac{\left|\vec{p}\right|^2(j)}{p^0(j)} \quad,
\end{equation}
and the chemical potentials are extracted by calculating ratios of the particle yields.  Extracting the entropy accurately from a microscopic 
transport model requires some thought; the medium cannot be simply treated as an ideal gas of massless particles, and one should note that the specific entropy contributions from particles of different masses are not the same.  We verify our calculation for s via a scaling relation between the entropy density and the temperature of an equilibriated system at fixed volume: $s\sim T^{\frac{1}{c_s^2}}$, with $c_s$ the speed of sound. The left frame of Fig. 2 shows pressure as a function of energy density. 
%for a sample at fixed ground state nuclear density.  
The speed of sound can be extracted from $c_s^2=\left(\frac{\partial P}{\partial \epsilon}\right)$, evaluated along an adiabat; the slope from the left frame in Fig. 2 yields $c_s^2=0.18$.  This value has been fitted to the extracted entropy densities as a function of temperature and hence satisfies the entropy scaling relation.  In addition to calculating the entropy via the Gibbs formula, we have calculated the entropy by determining the individual particle species multiplicities in the system and summing over them, weighted with their specific entropies (such specific entropies are extracted via the aforementioned statistical model).  Since both methods for the entropy extraction agree to within 5-10\%,  and the entropy scaling relation is verified, we conclude that our entropy computation is accurate and represents a hadronic medium including multiple different particle species.

\begin{figure}[tb]
\includegraphics[width=0.9\linewidth]{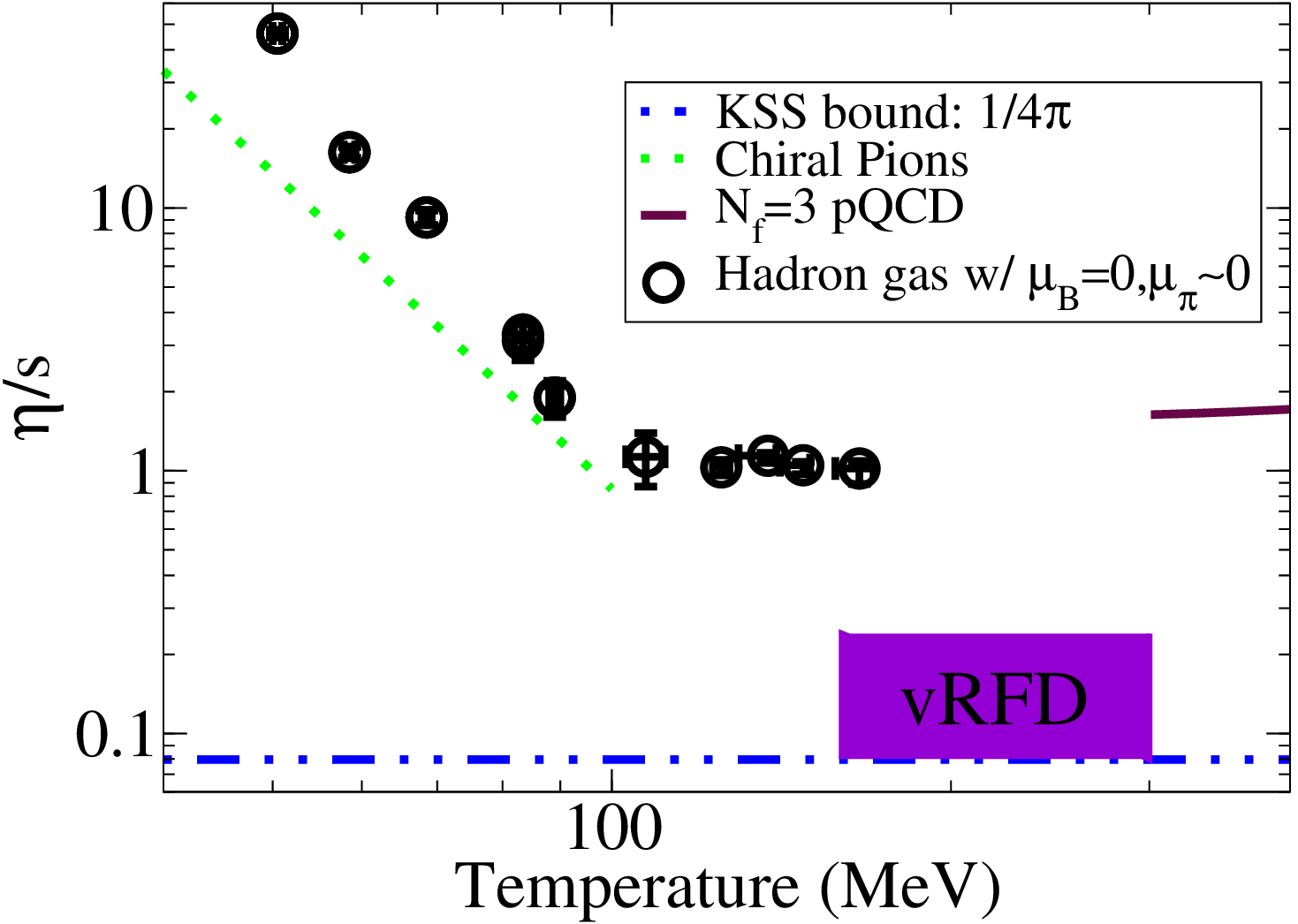}
\caption{The viscosity to entropy ratio in chemical and kinetic equilibrium.  Also shown is the result of $\eta/s$ for chiral pions, and a pQGP of 3 massless quarks.}
\label{Fig-eta_s_zmu}
\end{figure}

Using the aforementioned techniques, we present $\eta/s$ as a function of temperature in 
full equilibrium for zero baryo-chemical potential in Fig. 3. Also illustrated in that figure is the calculation of $\eta/s$ for chiral pions \cite{Prakash:1993bt} and 3 flavor perturbative QCD \cite{Arnold:2003zc}. Our results are in qualitative agreement with the result for chiral pions -- they suggest that the minimum value of $\eta/s$ in a hadron gas should occur near $T_c$ \cite{Csernai:2006zz}.  The minimum value found for $\eta/s$ for the equilibrium and zero 
baryo-chemical potential case is $\approx 0.9$, significantly higher than the KSS bound of $\eta/s \approx 0.08$.  If the minimum value of $\eta/s$ for hadronic matter in the range of hadronic freezeout indeed occurs at that value, our results of $\eta/s$ would pose a serious problem for the application of viscous hydrodynamics to the hadronic collision evolution at RHIC, since a value of $\eta/s$ of at most 0.24 is needed to reproduce RHIC elliptic flow data \cite{Romatschke:2007mq}.

However, calculating $\eta/s$ in full kinetic and chemical equilibrium, as has so far been common practice, may be unsatisfactory: while a statistical model analysis of particle yields and ratios at RHIC indicate a chemical freeze-out temperature
in the vicinity of $T_{chem}\approx~160$~MeV \cite{BraunMunzinger:2001ip,BraunMunzinger:2003zz}, hydrodynamic calculations indicate that a kinetic freeze-out temperature of $T_{kin}\approx~130$~MeV is required to describe the momentum distributions of final state hadrons \cite{Hirano:2002hv,kolb:2002ve,Kolb:2003dz,Nonaka:2006yn}. The separation of time-scales (and hence temperatures) for chemical and kinetic freeze-out imply that the hadronic phase of a relativistic heavy ion reaction actually acquires increasingly non-unit fugacities as the system cools after chemical freezeout and evolves out of equilibrium \cite{Hirano:2002hv,kolb:2002ve}. One should note that microscopic calculations \cite{Bass:2000ib,Nonaka:2006yn} of the hadronic evolution do not require explicit introduction of such fugacities, since the respective phenomena are the result of dynamically changing collision rates for inelastic vs. elastic hadronic rescattering processes.  In the context of our hadron gas calculations, a non-unit fugacity can be induced by initializing the system with a surplus of pions (kaons) relative to the 
$\lambda_{\pi,K}= \exp(\mu_{\pi,K}/T)=1$ value for the corresponding temperature and performing the viscosity measurement before the system relaxes into chemical equilibrium.  A similar effect can be obtained by inducing a finite net-baryon density, which results in a finite baryon-chemical potential.

Figure 4 displays our calculation for $\eta/s$ as a function of temperature at non-unit fugacities or at finite baryo-chemical potential. 
%We find that our results for finite chemical potential(s) differ depending upon how the chemical potential compares to the temperature in the system.  
The largest effect is seen for finite baryochemical potentials.  The reduction in $\eta/s$ can be understood classically if one associates a non-unit fugacity with an increasing particle density of that corresponding species in the system.  Since $\eta \sim \frac{\bar{p}}{\sigma}$ (with $\bar{p}$ the mean momentum of the particle), increasing the particle densities of a given species in the system will lead to a reduced mean free path, which in turn  reduces the viscosity (in the case of adding baryons, the average cross section will rise, with similar effect).  Similarly, increasing the multiplicities of different species in the system will enhance the entropy density, hence $\eta/s$ will decrease.  Taking non-unit fugacities into account, our values for $\eta/s$ in the range of hadronic chemical freezeout can be as low as $\eta/s \approx 0.4-0.5$.  However, this is still significantly above the suggested values $\eta/s \approx 0.08-0.24$ from viscous hydrodynamics calculations, and we should keep in mind that the hot hadronic matter created at RHIC, though not at unit pion or kaon fugacities, still is most likely at $\lambda_{B}\approx1$.  This observation enables us to constrain the origin of the low viscosity phase in a relativistic heavy ion reaction.  While a perturbative QGP is expected to have a large value of $\eta/s$, it should strongly decrease as a function of decreasing temperature in the strongly coupled non-perturbative region, and its minimum should be reached at or near $T_c$ on the deconfined side of $T_c$.  $\eta/s$ may then exhibit a sharp rise or even a discontinuity as it crosses $T_c$ from $T\simeq T_c^+$.  Discontinuities in $\eta/s$ at $T_c$ have been computed in weakly-coupled scalar field theories, and arguments have been given to suggest the discontinuity in $\eta/s$ at a phase transition could be a universal feature of a much larger class of systems \cite{Chen:2007jq}.   

\begin{figure}[tb]
\includegraphics[width=0.9\linewidth]{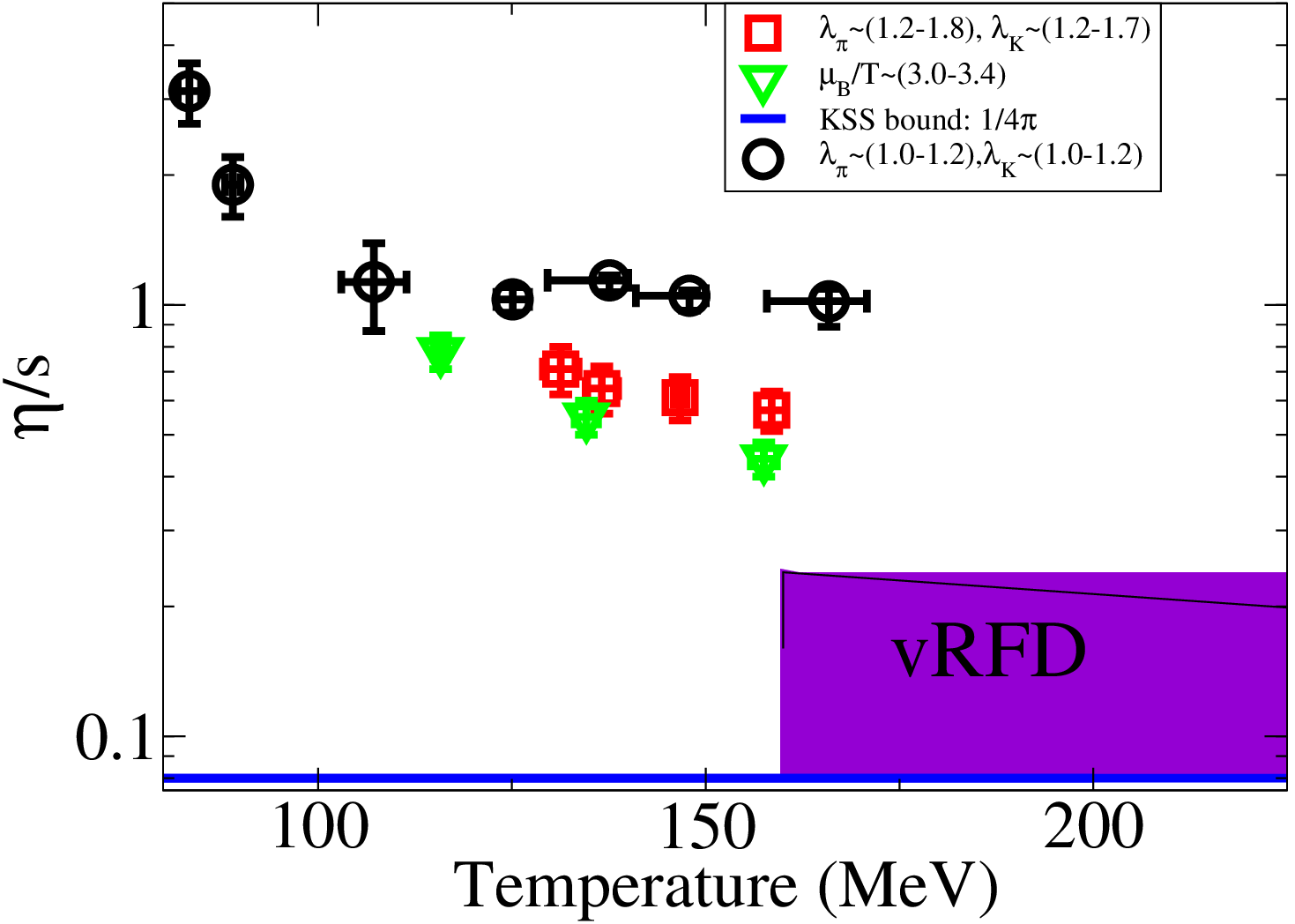}
\caption{The shear viscosity and viscosity to entropy ratio as a function of temperature for different values of $\frac{\mu_B}{T}$ and $\lambda_{\pi,K}$.  Note the strong decrease in $\frac{\eta}{s}$ for large values of the fugacities.}
\label{Fig-eta_s_fmu}
\end{figure}

In summary, we have calculated the viscosity over entropy density ratio $\eta/s$ of a hadron gas as a function of temperature, baryo-chemical potential and fugacities. We have demonstrated that the inclusion of non-unit particle fugacities, which are bound to arise due to the separation of chemical and kinetic freeze-out during the heavy ion collision evolution, will reduce the value of $\eta/s$, but not to the value necessary to ensure the successful application of viscous hydrodynamics to the full collision evolution at RHIC.  Our calculation of $\eta/s$ in a hadron gas from a microscopic transport model therefore constrains the origin of the low viscosity matter produced in a relativistic heavy ion collision, which must occur in the deconfined phase, possibly in the range $1 < T/T_c < 2$. At the formation of the hadronic phase, which is thought to occur in chemical equilibrium, $\eta/s$ will experience a sharp increase. However, subsequently its value may decrease again due to the system evolving out of chemical equilibrium. Near kinetic freeze-out $\eta/s$ will rise with decreasing temperature.  This lends credence to the notion that the dynamics of the evolution of a collision at RHIC is dominated by the deconfined phase exhibiting very low values of $\eta/s$.

\acknowledgements
This work was supported by DOE grants DE-FG02-03ER41239 and DE-FG02-05ER41367.
We wish to thank Berndt M\"uller, Ulrich Heinz, Jorge Casalderrey-Solana, Derek Teaney, Giorgio Torrieri, and Pasi Huovinen for helpful discu
ssions and suggestions.  
%\references
%\bibliography{/users/bass/Publications/SABrefs.bib}
\bibliography{hadronic_etas4.bib}

\end{document}